\documentclass[conference, twocolumn]{IEEEtran}
\usepackage{mdwlist}
\usepackage{cite}

\usepackage{amsmath}
\usepackage{amssymb}
\usepackage{amsthm}

\usepackage[final]{graphicx}
\usepackage{color}
\usepackage{subfigure}
\usepackage{epstopdf}

\newtheorem{theorem}{Theorem}[section]
\newtheorem{proposition}[theorem]{Proposition}

\theoremstyle{definition}

\theoremstyle{remark}
\newtheorem{remark}[theorem]{Remark}

\title{ An Informed Model of  Personal Information Release in Social Networking Sites}
\author{
\IEEEauthorblockN{Anna Squicciarini}
\IEEEauthorblockA{College of Information Science and Technology\\
Penn State University\\
University Park, PA 16802\\
E-mail: \texttt{asquicciarini@psu.edu}}
\and
\IEEEauthorblockN{Christopher Griffin}
\IEEEauthorblockA{Applied Research Laboratory\\
Penn State University\\
University Park, PA 16802\\
E-mail: \texttt{griffinch@ieee.org}}
}

\begin{document}
\maketitle

 \begin{abstract}
 The emergence of online social networks and the growing popularity of digital communication has resulted in an increasingly amount  of information about individuals  available on the Internet.  Social network users  are given the freedom to create complex digital identities, and enrich them  with  truthful or even   fake  personal information. 
 However, this freedom has led to serious security and privacy incidents, due to the role users' identities play in establishing  social and privacy settings.

 In this paper, we  take a step toward a better understanding of online information exposure. Based on the detailed analysis of  a sample of real-world data, we   develop a deception model for online users. The model uses a game theoretic approach to characterizing a user's willingness to release, withhold or lie about information depending on the behavior of individuals within the user's circle of friends. In the model, we take into account both the heterogeneous nature of users and their different attitudes, as well as the different types of information they   may expose online. 

  \end{abstract}
\section{Introduction}
Online social networks (OSNs) such as Facebook, Myspace, and Google+  allow individuals to present themselves and establish or maintain connections with 
others. Users articulate their social networks by  creating and managing content, social connections, and a possibly large amount of personal information. A typical OSN in fact allows users to create connections to ÔfriendsÕ, thereby sharing with them a wide variety of personal information. These connections  are often based on the alleged identities and properties of  the individuals populating the OSN.    
  Users of social media sites can, however,   generate accounts containing unverified information. On the one hand, this allows the users   to avoid identification and surveillance or observation of any kind. 
 On the other one,   the ability to generate unverified accounts on most of these sites, renders social relationships potentially weak, if based on fake identities. Further, unverified accounts may and are often used by malicious users to carry out disruptive activities hidden behind fake identities \cite{donath2011}.   
%
To date,  while some work has studied the incentives behind information disclosures in OSNs \cite{idm1,idm2, Lampe:2007}, little is known about  identities misrepresentations. 

  In this paper, we speculate that information revelation in OSN  is a complex process where multiple  contrasting influences are in play:  not only privacy attitudes, but also social pressure and
  personal attitudes are at stake. 
Focusing  on three types of users' behavior related to information revelation:  truthful information sharing,   information withholding and deception, we  study the effect of misrepresentation in these environments by means of a game theoretical model.

To ground our model, we conducted an extensive empirical study,  collecting data about users' common behavior  and their attitude toward personal information disclosure. The study involved  almost 300 subjects, all active social network users.

Our study reveals important insights on users' attitudes and practices. In particular, our results show that users'  decisions to lie or withhold information
are not strongly influenced by privacy concerns. Rather, results show strong correlations between peer-pressure and attitudes toward lying. Users who feel peer-pressured are less likely  to  withhold  their information, especially their whereabouts. The quest for gaining or maintaining popularity also seems to play an important role, in particular with respect to the amount of information users choose to reveal.   Also, we identified that users' identity information is managed differently depending upon the perceived sensitivity of the information. For example, for basic demographic information, users tend not to lie in the main social network account, as this is typically revealed in the course of social interactions and may be easy to verify by social network peers. On the other hand, information that is closer to the users' personal sphere, for example, social relationships, whereabouts, etc. is  revealed mostly by users who are in search or popularity and/or are searching for self-affirmation in the network.   In addition,  we find that misrepresentation interacts with measures of morality, suggesting that users do not associate lies in social networks with unethical behavior,  and that, where lying is considered unethical, they are more likely to withhold information, as a form of boundary control.   Finally, we found that users' behavior is mostly influenced by inner circles of close  online friends,  regardless of the actual number of social connections users have. 

The analysis of the responses is used as input to inform our  qualitative model of user information sharing, withholding and deception.  In particular, building on the finding that users treat information differently, the model presupposes that individuals release, withhold or lie about certain classes of information differently, and that each user  behaves according to an individual payoff function. The payoff function is constructed to take into account the identified influential decision factors: morality, peer pressure, privacy and popularity. The output of the function is also affected by the behavior of a circle of close friends - as we found strong evidence of self-validation and peer influence in our study.
 We provide an example model using evolutionary dynamics, which we posit influences a users' behavior as he interacts with his OSN overtime and more accurately understands the true nature of his (personal) objective function.

The paper is organized as follows. Next section reviews relevant literature. Section \ref{sec:III} discusses  our empirical study. 
  Section \ref{sec:model} presents our model.
  We illustrate the various types of users and information in our examples in Section \ref{sec:EvolutionaryGame}. We conclude the paper in Section \ref{sec:conc} with pointers to future research directions. 

    \section{Literature review}
  Digital identity constitutes one of the building blocks of  Web 2.0 technologies, ranging from social networking to e-commerce.  Problems related to digital identity management and protection have been tackled by both the computer science community and by information scientists.  From the computational standpoint,   a  variety of digital identity and trust management mechanisms have been developed to allow users to create and maintain complex digital personas  \cite{gail,idm1}  although there has  been little  work on the topic   of digital identity validation and trust in the context of social computing.

From a social science perspective, various studies have explored identity sharing behavior in social network sites and the
risk of over exposure (notable examples are \cite{idm1,idm2,idm3, Lampe:2007}).  
  Research studies have shown that  users in
online environments rely on a variety of cues to make
determinations about one another; however, all these cues
are not deemed equally credible. For instance, Goffman
\cite{Goff} notes that identity cues can be intentionally given or
unintentionally  revealed, and that humans are more likely to
place greater weight on those cues that are perceived to be unintentional
as opposed to strategically constructed. This ability to
engage in deceptive self-presentation online is compounded
when  users do not share a social network and
therefore have less access to Òinformation trianglesÓ such as
mutual friends who might confirm or deny information.
Donath \cite{donath2011} argues that a shared social
network can provide explicit or implicit verification of
identity claims. Therefore, as highlighted in \cite{Lampe:2007}, a  highly connected network such as Facebook should
encourage more truthful profiles, or misrepresentations that
are playful or ironic as opposed to being intentionally
deceitful.
 
Burke et al. \cite{Burke} studied user motivations for contributing in social networking sites, based on server log data from
Facebook. They found that newcomers who see their friends
contributing go on to share more content themselves. 
Furthermore, those who were initially inclined to contribute,
receiving feedback and having a wide audience, were also
predictors of increased sharing.

    Complementary to the body of work on   identification and information revelation,  is the  work on anonymity in social network sites \cite{danfeng1,ano3,ano2}. The emphasis  in these works is however  on algorithmic approaches for  non-disclosure and anonymity preservation, rather than on actual revelation.
Finally, parallel to this body of work is the work on reputation \cite{rep,repLada}. Reputation of digital identities and trust in online environments have  been  investigated by multiple research communities ranging from computer science \cite{nurmi} to economics \cite{reput,repec}. 

With respect to our methodology, our work  employs analytical models. Analytical models for various security topics based on game, information and decision theories are rapidly growing in interest \cite{152}. In particular,  game theoretic approaches to reputation and trust first emerged in the economics literature  
(a typical example is \cite{64}) and were then applied to online settings \cite{xx,KL,gamet}.   However,  to the best of our knowledge, the only work analyzing social identities using analytical tools  is from   Alpcan and colleagues  \cite{pain}.  Alpcan's work  focuses  on reputation and trust,  where strategies are defined in terms of opinion, quantified through a simple cost function.  As we discuss in the next sections, our focus is on validation and individual attitudes toward deception, rather than lies.     
    Additionally, an interesting economically inspired work dealing with users' privacy is discussed by Papadimitrous and colleagues \cite{KL},  who propose a precise estimate of the value of the private information disclosed by a set of individuals, and a compensation for such information release that may induce users to release richer information. Yet, the model applies to a different set of applications, such as online surveys and e-commerce applications.
    
    This work is part of our research effort on deception and information revelation in social networking sites.  In \cite{SSG11} we studied the interaction of users and servers at the time of user registration, and used  a game theoretical
framework to describe a simple two-player general sum game
describing the behavior of a server system (like Facebook) that
provides utility to user. We  showed that in
the presence of a binding agreement to cooperate, most players
will agree to share information. In \cite{GS12}, we investigated a simpler game model in which rewards for releasing information and costs for withholding information and lying were represented by arbitrarily chosen concave and convex functions. We showed for a specific instance of a payoff function that a symmetric Nash equilibria existed and was related to the automorphism class of the graph describing the interaction graph of the social network. This work substantially extends our previous work by more accurately modeling the qualitative nature of the user's objective function through the incorporation of information in our survey.  Our previous work was purely theoretical, and used a overly simplified the notion of identity.  Identity  was mainly considered as an atomic value, and  therefore  was focused on different aspects of information sharing in social networking sites (for example the registration of new users). We also incorporate a model of evolutionary dynamics to explain a user's choices as he interacts with his social network and is exposed to his friend's choices.

\section{Informing the model through an exploratory study}\label{sec:III}

In order to understand  typical social network users' attitudes and actions with respect to information disclosure, we conducted an exploratory study using real-world data.  
The specific aim of our   study was to gain a deeper understanding of usersÕ identity-revealing actions, the peculiar features of average users, and the perceived understanding of identity on social sites.   

\subsection{Methods}
We conducted  a web-based  survey, collecting a total of 296 responses.  Respondents were recruited from two different undergraduate courses in the college of Information Science and Technology at the Pennsylvania State University.  One extra credit point for the course was awarded for their participation in the study.  
The survey was constructed to study  three specific aspects of users' behavior: (1) privacy awareness, (2) attitude toward information withholding and practices (3) attitude toward lies and misrepresentation.
 
 The respondents were aged between 20 and 35 ($\mu$=23, sd=2.34). 
The respondents were 65\% male and 35\%  female.  99.3\% of them declared to have at least one account on social sites, and 12\% declared to have more than one account on the same OSN. 
Participants were asked to indicate the social network they
most often accessed: 95.3\% most often accessed Facebook, while the remaining
participants were distributed among Google+, Linkedin (6\%) and Twitter. 
In terms
of network usage frequency, 94\% of the respondents accessed social network sites
at least once a week, and 83.6\% of those were daily users.

Considering that
Facebook is one of the social networks that most heavily promotes  personal information disclosure, our sample was deemed appropriate for this study. 
While the overall sample reflects a specific subset of the population, we notice that most of the active users in Facebook, according to recent statistics, are below 26 years old (and specifically in the 21-24  age range) \footnote{http://www.socialbakers.com/facebook-statistics}.

 The instrument also included five broad types of measures of  perceived privacy, social pressure, and popularity (or social capital), which serve as dependent variables.


\subsection{Measures}
\begin{itemize}
\item {\em Deception}  was our independent  measure, and was  measured by two sets of 4 items each. The first set focused on deceptive activities, and was measured on a  a frequency rating scale
(1=all of the time to 5=never).  The second set of items related to the perception about deception (lies and withholding information on social networking sites). An example item is ``Lying  in social network is 
unethical''.  These items were also rate using a Likert scale (5-point rating scale, where 1= strongly agree
and 5=strongly disagree).
\item{\em Usage} was measured using 6 different items. Three of the items where focused on frequency of usage and number of connections. The remaining items where added  to ascertain the extent to which the participant engages in certain 
types of social interactions,  e.g., posting images, giving feedback to other's posts or images, sharing a url, tagging a video or an image.  For these items, we used  a frequency rating scale
(1=never to 5=once or a few times a day)
\item {\em Privacy Concerns.}
Individual differences in privacy perceptions  can be significant  \cite{XuDSH08, YaoRW07}.  Thus, we need to establish a baseline understanding about the awareness of and attitudes toward privacy protection by participants.  In our survey, we included five questions to ask participants about their information disclosure behaviors and privacy concerns in Social Networking sites (Cronbach $\alpha=.71)$, rated on a Likert scale (5-point rating scale, where 1= strongly agree
and 5=strongly disagree). An example item is "I have had concerns about the privacy of my data on Social Networks".  
\item {\em Pressure} was measured using 5 items (Cronbach $\alpha=.823$), focusing on pressure of updating information (e.g. ``I feel peer pressured to constantly update my Social Network profile") and uploading content. The items measured perceived pressure from the social networking and  from peers (e.g. ``I need to update my profile often to be popular among my friends").
\item {\em Perceived Popularity}  was measured by 6 items (Cronbach $\alpha=.732$), focusing on the impact on one's popularity (or social capital) upon passively being involved in  the   social interactions listed in the usage measures.
\end{itemize}

\begin{table}[t!]
\begin{center}
\begin{tabular}{|c|p{2.6cm}|c|c|}
  \hline
{\bf ID} &{\bf Question} & {\bf Average} & {Standard Dev.} \\\hline   
Q1 & I have put false information in my main social network account (1=strongly agree, 5=strongly disagree)& 1.87 & 0.4 \\ \hline 
Q2 & I have withheld information from my main social network account (1=strongly agree, 5=strongly disagree) & 1.91& 0.732\\ \hline
Q3 & Putting false information about myself and my whereabouts on my profile can help me be more popular	& 2.45& 0.453\\ \hline
\end{tabular}
\end{center}
\caption{\small Descriptive Variables concerning Deceptive activities}  \label{tab:1}
\end{table}

\subsection{Findings} \label{sec:find}
The main purpose of this study was to examine usersÕ attitudes towards
deception  in social networking sites.  

  Leveraging results from  previous research studies \cite{donath2011, Lampe:2007}, we hypothesized that  (h1)  users in fact deceive in social networking sites, but  mostly choose to portray truthful  portions of their  basic identity,  which could be validated offline (e.g. the name or gender),   and deceive or withhold  information  which may be deemed too personal or inappropriate for disclosure to the social network audience.   
 We also hypothesized that participants decision  to withhold information or lie would be influenced by  (h2)   their privacy inclination, (h3) their perceived pressure to be active on the social network site  and (h4)  their
 wish to be popular  among peers.   
Finally, we were interested in learning whether users'  decision to withhold or lie would be connected with ethical choices.  Here, we did not have an initial hypothesis, but were interested in exploring the correlations between morality and deception.
We discuss our results in a detailed manner in the following. 
\vspace{3pt}

\subsubsection{\bf h1:  Frequency of Deception} \label{infotype}
\begin{figure}[t!]
\begin{center}
\includegraphics[width=0.4\textwidth]{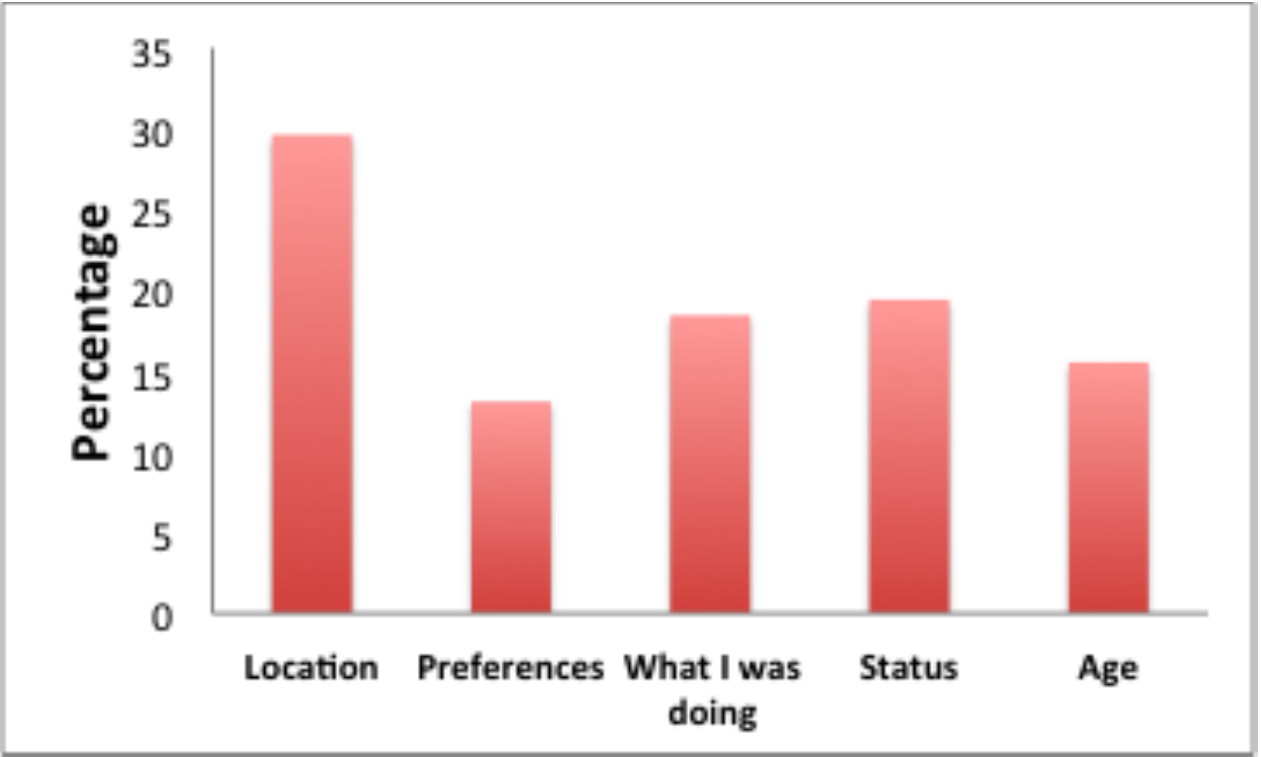}
\caption{Data Items most frequently misrepresented}
\label{fig:3}
\end{center}
\end{figure}

We began with identifying whether deception is in fact  significant. 
Table \ref{tab:1} presents the descriptive statistics
of some of  the study's variables related to deception.   As reported,   a vast majority of the participants admit to having lied at least once, and also chose to withhold information (94\% of respondents either agree or strongly agreed to have lied -the exact statistics are reported in the table).   Figure \ref{fig:3} highlights the specific types of attributes users most often lie about. 
 We further  determined that users who are more likely to be involved in discussions and are therefore active in the social networking site
report a lower frequency of lying (Pearson r=0.277, p=0.033), therefore reinforcing the well-known   signaling theory identified by Donath \cite{donath2011}. The relationship with the ``withholding question'' shows a similar trend, but it is not statistically significant, therefore a conclusive statement on this relationship is not possible. 
Users also report it is easy to detect  lies of  their close social connections with whom they often interact with ($\mu=2.61$, sd. 0.912), again confirming that self-validation is effective in social networking sites. 

\subsubsection{\bf h1: Types of Information Revealed}  To further explore which pieces of information users are likely  to withhold or lie about, we asked users to indicate their preferred action  for six different personal pieces of information: location, gender, GPA, relationship status, telephone number, current occupation. We select properties that would be considered important and potentially sensitive for our participants, who were mostly students.  Users were given the option to indicate for each attribute one of three choices: tell the truth,  provide false information, do not put anything. 

\begin{figure}[t!]
\begin{center}
\includegraphics[width=0.5\textwidth]{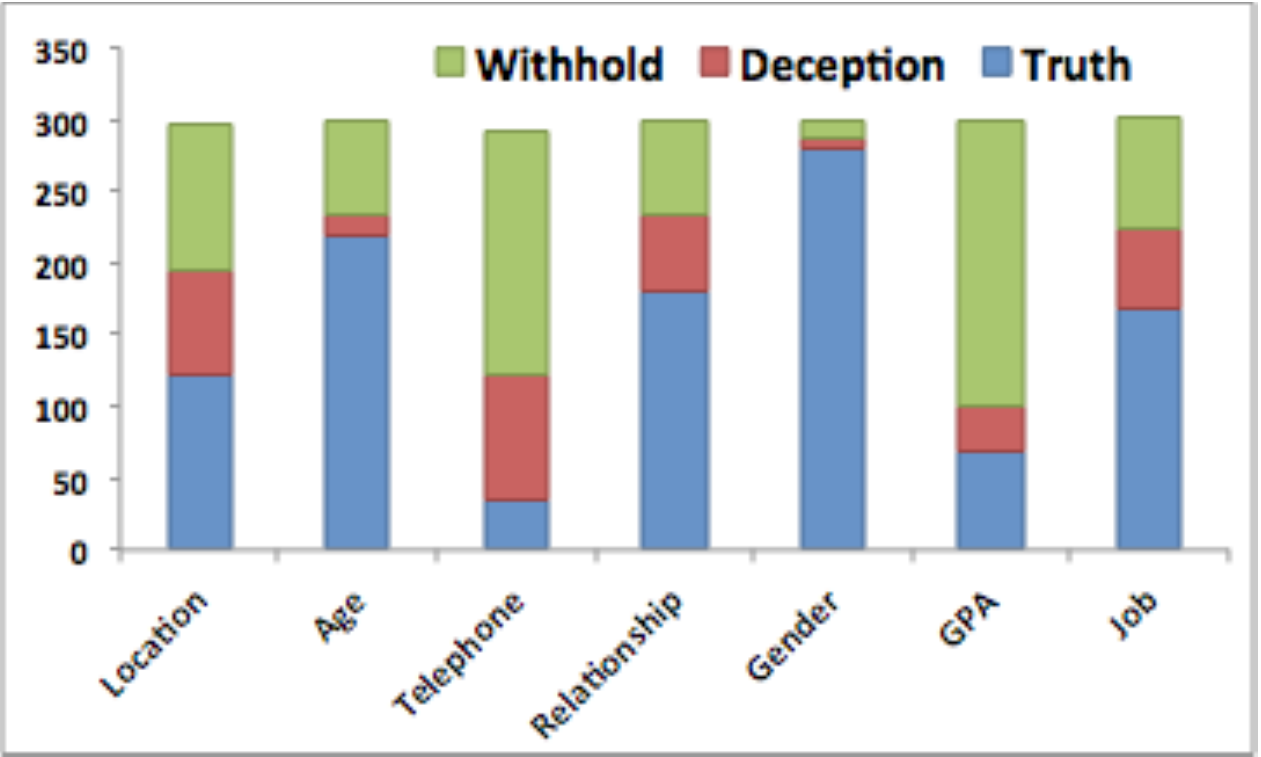}
\caption{Responses breakdown by attribute}
\label{fig:break}
\end{center}
\end{figure}

The responses, organized by attribute, are reported in Figure \ref{fig:break}. 
In our survey, most of the participants claimed to misrepresent only specific pieces of information. In particular, our analysis confirms that highly interconnected users are likely to reveal basic identity properties, such as gender, age, etc. truthfully (Pearson .436, r=0.012).  
 Information commonly deemed as private, such as telephone number and GPA, is instead  mostly withheld, or misrepresented.  Finally, there is some interesting variability with respect to  location, current occupation and  relationship status, where there is not a  predominant choice.
These results  confirm our  hypothesis (h1). 

 \vspace{3pt}
\subsubsection{\bf Influential factors of information sharing} \label{factors}
The analysis of the factors influencing information sharing activities resulted in the following findings.  
\begin{itemize}
\item {\em h2: Privacy}  We  first  analyzed the responses related to {\em privacy awareness}, to get a sense of the respondents attitude toward information revelation and leakage in social networking sites. 
 An initial notable result is that,   despite the fact that most respondents maintained a detailed profile on their favorite social network site, many of them also demonstrated relatively high levels of privacy concern.  The responses to the statement ``I maintain a detailed profile on my main social network account'' confirm that they maintain rich profiles  ($\mu$=1.97,  std=0.96),  and that they reveal their main identity for the most part ($\mu$=2.01, std=0.45).    Nevertheless, their responses to the statement ``There is a high potential for loss involved in sharing personal information on Social Networks like Facebook'' indicate their awareness of potential information leakage ($\mu=1.80$, std=0.81).  
  We then tested our first hypothesis, i.e. whether lying or withholding information was related at all to the respondents level of privacy awareness. We conducted an exploratory least-squares multiple regression
analysis, regressing  their responses to question Q1,  with  their responses related to privacy concerns as predictors.  None of these appeared to be strongly related.
  
\hspace*{1em} The results lead to interesting findings.  First, in general participants are concerned with their privacy on social networking sites and are aware of the potential loss of privacy;  second, the results confirm the existence of phenomenon known as the Òprivacy paradoxÓ  \cite{xu}, in which individuals state that they have privacy concerns, but behave in ways that seemingly contradict these statements by providing detailed information about themselves.  Finally, our results show that  privacy is not indicative of their choice to deceive, or withhold information.

\item {\em h3: Pressure} Next, we investigated  whether {\em participants  feeling   peer pressured   are more likely to  deceive or withhold information}. 
We first tested whether feeling  peer pressured would be correlated with the amount of personal information displayed on the social network site. We conducted a  simple regression analysis,  using the answer to the question ``I feel peer pressured to constantly update my profile'' as an independent variable, and their self-declared level of detailed  social network profile as a dependent variable.  The test shows that the more users agree to feeling pressured to update their profile, the more they claim to display a detailed profile (Pearson r=0.433, p=.034). 
We  then studied  whether the information being revealed upon being pressured is truthful or not. We  correlated Q1 and  Q2 with  our items related to popularity. 
Our results  show that there is not a significant correlation between users' perceived peer-pressure and their choice to deceive. 
However, there is a clear correlation between their choice to withhold information and their feeling of being peer pressured (Pearson r=0.163, p=0.03). That is to say, the more users feel peer pressured, the less likely they are to withhold information. 

\item {\em h4: Popularity}  When correlated with measures relative to {\em popularity},  we obtained the following results.  First,  we correlated  participants'   frequency  of sharing content (e.g., images) in the social networking site  with their perceived popularity gain by doing so. We   obtained a significant correlation  (Pearsons r=0.272, p=.032). In line with previous studies in this space \cite{Lampe:2007},   this finding  shows that the more users perceive certain social interactions to benefit their social capital, the more likely they are to  pursue them.   With respect to deception, the majority of participants  disagreed to have  lied to gain popularity or portray a different  ``self" (only 25\% of respondents either agreed or strongly agreed to the question "I have put false information to appear different from my original self"). However,  we discovered a significant correlation between their   quest for popularity
 and their deceptive activities (Pearson r=-0.2449, r=0.015), therefore confirming our hypothesis. 
 
  \end{itemize}
  
   \begin{table}[t!]
\begin{center}
\begin{tabular}{|c|p{3cm}|c|c|}
  \hline
{\bf ID} &{\bf Question} &   {\bf Avg} & {\bf St. Dev.} \\\hline   
Q4 & I consider lying in social network sites unethical & 3.30 & 0.943 \\ \hline
Q5 & I consider withholding information in social network sites unethical  &4.10& 0.842 \\ \hline
\end{tabular}
\end{center}
\caption{\small Morality and Deception (Likert Scale: 1=strongly agree, 5=strongly disagree)}  
\end{table}
In summary,  this study  confirms that a social network user's tendency to deceive for certain data types  is highly correlated with his or her desire to portray a successful Òsocial image,Ó and not statistically related to privacy concerns. In other terms, the perceived usefulness of the social network service increases online usersÕ willingness to disclose their personal information. 

 \vspace{3pt}

\begin{figure}[t!]
\begin{center}
\includegraphics[width=0.3\textwidth]{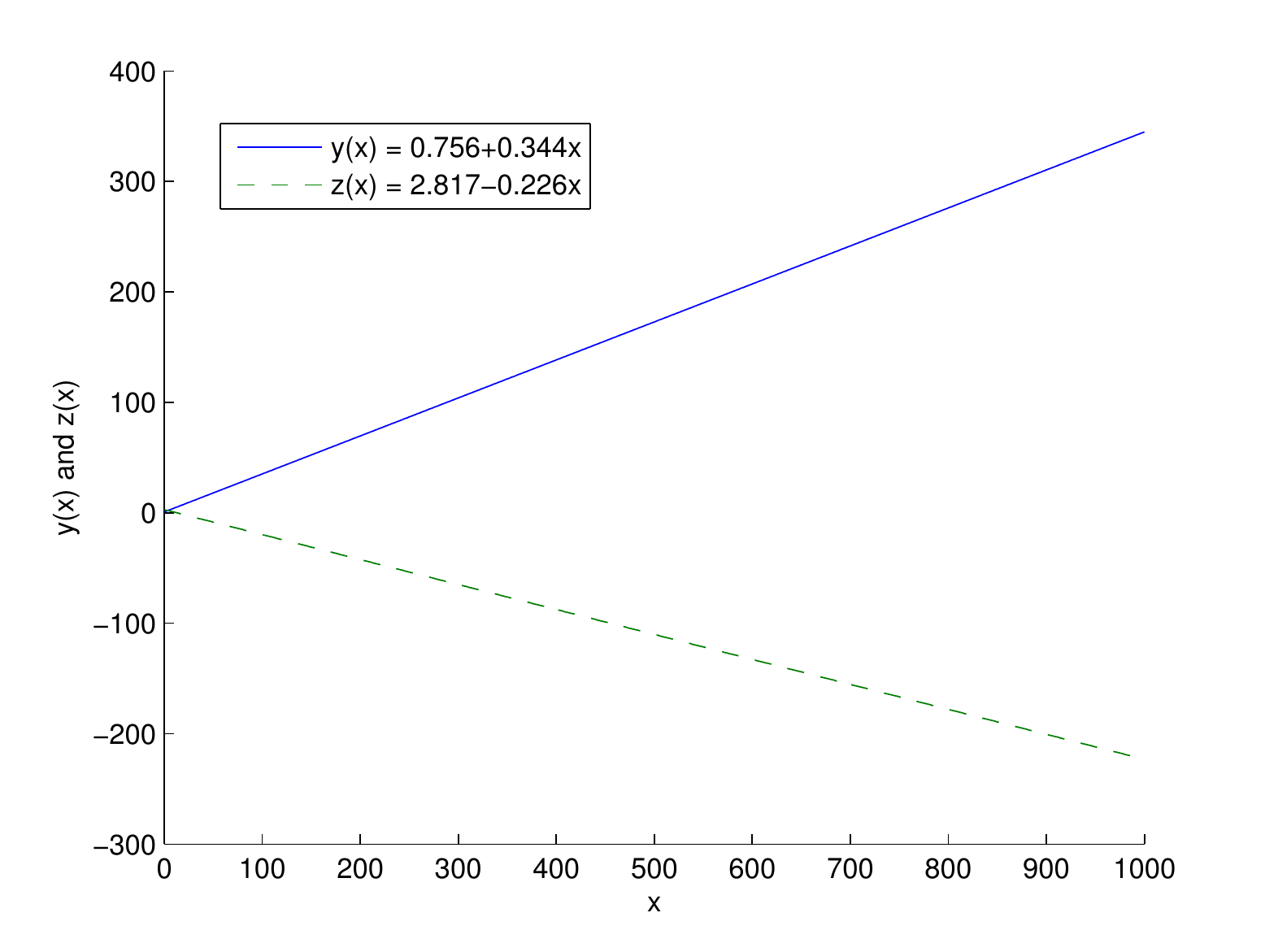}
\caption{Linear regression for questions correlating Q1 and Q4 (z(x)) and Q1 and Q5 (y(x))}
\label{fig:lreg}
\end{center}
\end{figure}  
%

\subsubsection {\bf  Morality in Social Networking Sites} \label{sec:moral}
 Our results show   non-obvious relation between lying or withholding information on a social networking site  and  morality. 

The results of correlating social network lying (corresponding to Q1) with  Q4 and Q5 are interesting, as they show opposite effects of thinking that lying on a social network is unethical and withholding information on a social network is unethical.  A higher value for Q4 (meaning an individual disagrees that lying is unethical) predicts a higher frequency of putting up false information on a social networking site (Q1).  However, a higher value for the withholding information questions (Q5)  predicts a lower frequency of putting up false information. The linear equations obtained through regression analysis are reported in Figure \ref{fig:lreg}.   This result seems to reinforce the notion  that lying on a social networking website and withholding information function as two completely different actions and that a user will choose one or the other based on an internal utility function.

  \begin{table}[b!]
\begin{center}

\begin{tabular}{|c|p{2.2cm}|p{2.4cm}|c|c|}
  \hline
{\bf ID} &{\bf Question} & Scale &  {\bf Avg} & {\bf St. Dev.} \\\hline   
Q7 & How many friends do you think typically check your profile &  1=``$> $150", 2=[100,150],
3=[50,100],
 4=$<$50 &3.63 & 0.658 \\ \hline 
Q8 & How many friends do you check typically & 1=``$> $150", 2=[100,150],
3=[50,100],
 4=$<$50  &3.53& 0.760\\ \hline
Q9 & How many social connections do you have? &  1=``$> $150", 2=[100,150],
3=[50,100],
 4=``$<$50''   &1.48& .901 \\ \hline
Q10 & How often do you visit your favorite social network site?&1=Once or a few times a day		
2=Once or a few times a week	
3= Once or a few times a month	
4= Never &1.21& .464 \\ \hline
\end{tabular}
\end{center}
\caption{\small Social network usage}  \label{tbl:usage}
\end{table}

\subsubsection {\bf Inner  Circles}   Some other interesting findings were related to the existence and importance users give to inner circles within their social network. Despite the complex social connections tying users together, users are most strongly influenced by a small set of connections with whom they interact regularly and whose opinion counts to them. By correlating Q7 and Q8 (Table \ref{tbl:usage}), we discovered that regardless of the number of social connections users have,  most users check and believe their profile is checked by a parallel  number of users (Pearson  r= .521, p$<$0.001).  Most of the actions (e.g., comments and feedback) users perform involve inner-circle users, who are the ones influencing usersÕ decisions about lying and not lying.


\section{Game Theoretic Model of User Behavior} \label{sec:model}
We build on  the game theoretic approach to modeling users' actions in a social network begun in \cite{SSG11, GS12} to qualitatively explain the behavior observed in our experimental results.

We have identified that users treat types of information differently with respect to whether they disclose, withhold or deceive. Furthermore, we know that the behavior of users is highly dependent on the behavior of a small group of their immediate network neighbors.

 Let $G = (V,E)$ be a user graph for a social network and suppose we have several classes of information $\mathcal{I} = \{1,\dots,m\}$. Let $x^{(j)}_i \in [0,1]$ be the proportion of information type $i$ that Player $j$ will release and let $y^{(j)}_i \in [0,1]$ be the proportion of information type $i$ about which Player $j$ withholds. Then $z^{(j)}_i \in [0,1]$ is the proportion of information type $i$ that  Player $j$ lies about. Then we have: \begin{equation}
x^{(j)}_i + y^{(j)}_i + z^{(j)}_i = 1
\end{equation}
Let:  
\begin{equation}
\bar{x}^{(j)}_i = \frac{1}{|N(j)|}\sum_{k \in N(j)}x^{(k)}_i
\label{eqn:Average}
\end{equation}
where $N(i)$ is the neighborhood of Player $j$ in $G$. We make similar definitions for $\bar{y}^{(j)}_i$ and $\bar{z}^{(j)}_i$. These are the average network level of releasing, withholding and lying about information type $i$. In the presence of popularity measures (with respect to a given user's circle of friends) Equation \ref{eqn:Average} can be modified to be a popularity weighted average with form:
\begin{displaymath}
\hat{x}^{(j)}_i = \frac{1}{\sum_{k \in N(j)} p_k}\sum_{k \in N(j)}p_k x^{(k)}_i
\end{displaymath}
Here $p_k$ is the popularity weight for player $k$.

\begin{figure}[htbp]
\centering
\includegraphics[scale=0.38]{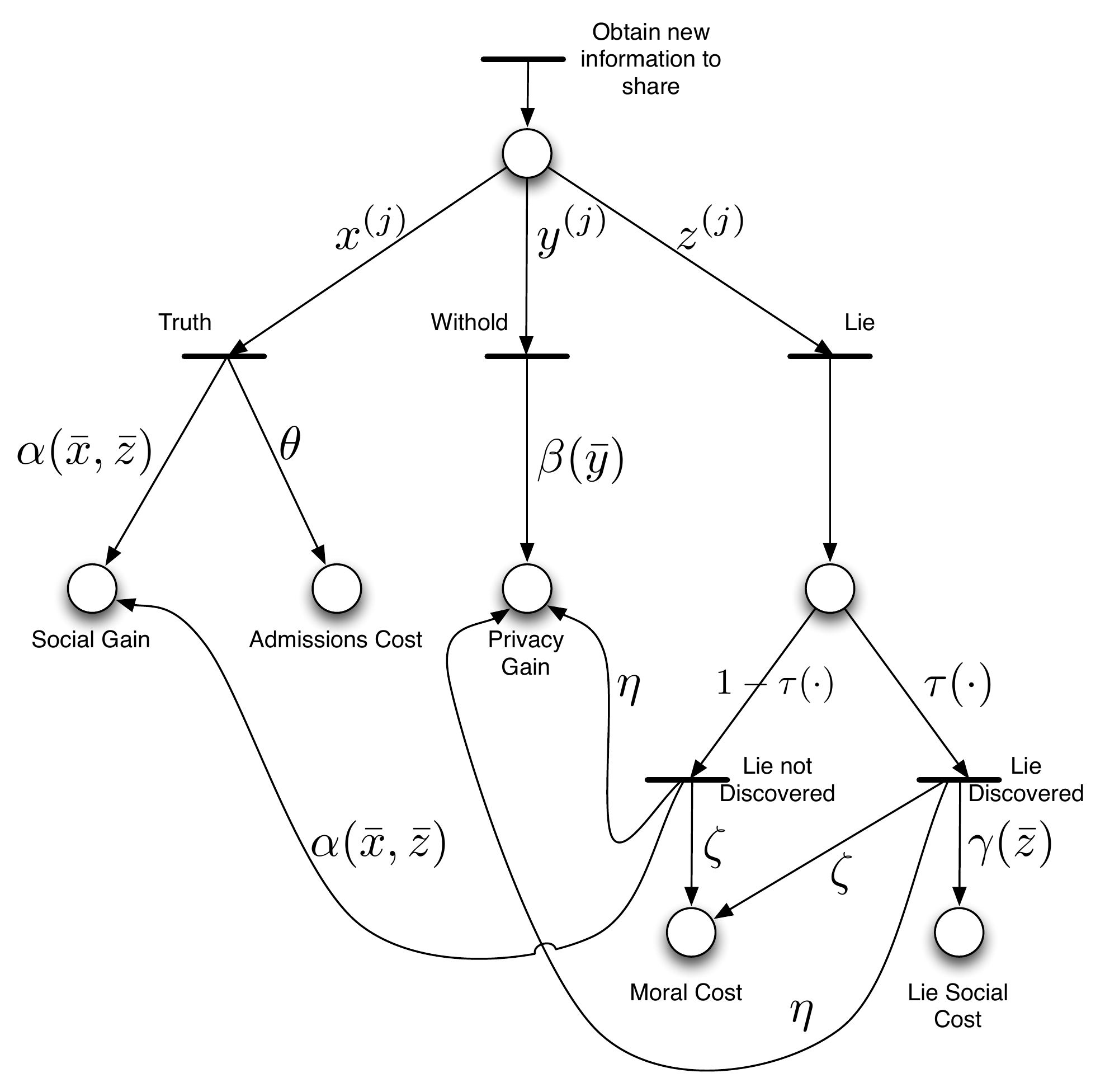}
\caption{A Petri net model of the accumulation of various components of the payoff associated to interacting in an online setting.}
\label{fig:PN}
\end{figure}

In \cite{GS12}, we assumed the existence of functions returning the reward for releasing information and costs for group deception and individual deception. We assumed these functions were concave, convex and convex (respectively), but provided no way to isolate their structure. We propose a richer model than the one in \cite{GS12}, which incorporates our observations from the empirical evaluation:
\begin{enumerate*}
\item The more users interact with the network, the less likely they are to deceive (see Sect. \ref{infotype}))
\item Users perceive interactions with their social network as a mechanism for gaining popularity, a form of social capital (see Sect.  \ref{factors}).
\end{enumerate*}
For the remainder of this section, assume that we've \textit{fixed} an information type (e.g. location, interest, age etc). Again leveraging our analysis (Section \ref{factors}), we assume that five elements make up each user's objective function:
\begin{enumerate*}
\item Social capital gained from sharing information within the group,
\item Personal benefit gained from maintaining information privacy,
\item Personal cost from the discovery of deceptive information,
\item Moral cost from deceiving a group and
\item The cost associated with admitting information (in exchange for social capital).
\end{enumerate*}
The easiest way to understand the relationship of these elements is as a token based model in which each action, causes a token (or fraction thereof) to be deposited into a specific revenue or cost bucket. The proposed model for this system is illustrated in the Petri net \cite{DA05} shown in Figure \ref{fig:PN}.  Given space constraints, we cannot formally define Petri nets. In short, they are graphical token models in which transitions move and spread within the vertices of a graph structure. The interested reader should see Chapter 1 of \cite{DA05}.\\
In general, we can think of the \texttt{Truth}, \texttt{Withhold} and \texttt{Lie} transitions as being controlled by the user with all other transitions being uncontrolled or controlled by nature. (Moody \cite{Moo98} discusses control in Petri nets.) Alternatively, as shown in Figure \ref{fig:PN}, we can think of the user controlling the fractional weights on the transitions leading to the \texttt{Truth}, \texttt{Withhold} and \texttt{Lie} transitions. If the Petri net is continuous, then we can think of the weightings leaving the controlled transitions as providing the benefit or cost for each token (or fraction thereof). We let $X^{(j)}$, $Y^{(j)}$ and $Z^{(j)}$ be correlated random variables whose dynamics are chosen by Player $j$. In general, $X^{(j)}$ is $1$ only if Player $j$ releases a piece of information, $Y^{(j)}$ is $1$ only if Player $j$ withholds information a piece of information and  $Z^{(j)}$ is $1$ only if Player $j$ deceives about a piece of information. Naturally, only one of these elements can be $1$ at any given time $t$ (thus we can think of these as being the outputs of a single discrete distribution) chosen by the player. At time $t$,  Player $j$'s stochastic payoff function is:
\begin{multline}
\Pi^{(j)}(t) =w_1\alpha(X^{(j)}(t),Z^{(j)}(t),\bar{x}^{(j)},\bar{z}^{(j)},t,\tau(t)) + \\
w_2\left(\beta(Y^{(j)},\bar{y}^{(j)},t)Y^{(j)}(t) + \eta(Z^{(j)}(t))\right) -\\
w_3\gamma(Z^{(j)},\bar{z}^{(j)},t,\tau(t)) - w_4\zeta(Z^{(j)}(t)) - \\
w_5\theta(X^{(i)}(t))
\label{eqn:SinglePayOff}
\end{multline}
In this expression:
\begin{itemize*}
\item $\alpha(X^{(j)},Z^{(j)},\bar{x}^{(j)},\bar{z}^{(j)},t)$ is a social capital function that provides the reward obtained by releasing a piece of information (true or false).
\item $\beta(Y^{(j)},\bar{y}^{(j)},t)$ is privacy capital function that provides the reward obtained by keeping a piece of information private.
\item $\gamma(Z^{(j)},\bar{z}^{(j)},t)$ is a cost function that yields the social price of lying about a piece of information. 
\item $\theta$ is an admissions cost function for each piece of true information revealed.
\item $\zeta$ is a moral cost function associated with each lie told.
\item $\eta$ is a privacy gain function associated to each lie (since a lie may protect privacy irrespective of any other moral judgement.
\item Finally $\tau(t)$ is the probability that a lie will be discovered by the social group. 
\end{itemize*}

Over a period of time, the complete stochastic payoff function for Player $j$ is:
\begin{equation}
\Pi^{(j)} = 
\sum_{t = 0}^{T}\rho^t\Pi^{(j)}(t)
\label{eqn:StochasticPayoff}
\end{equation}
The variables $w_i$ ($i = 1,\dots,5$) are the relative weights Player $j$ places on each component of his objective function. We can also think of $\tau$ as being a function of the total quantity of information (true and false) that has been released to the network:
\begin{equation}
Q^{(j)}(s) = \sum_{t = 0}^{s} X^{(j)}(t) + Z^{(j)}(t)
\label{eqn:Q}
\end{equation}
This provides consistency with two observations reported in Section \ref{sec:find}:
\begin{enumerate*}
\item Users who engage in their social network more frequently, tend to deceive less and
\item The more information available about a user, the easier it is for him to be trapped in a lie.
\end{enumerate*}
The parameter $\rho$ in Equation \ref{eqn:StochasticPayoff} is a discount factor chosen in the set $(0,1]$. It is worth noting that $\rho$ is only important if we wish to consider the limiting dynamics as $T \rightarrow \infty$. When $\rho < 1$, the user recognizes that future rewards have less value than rewards more immediately. To relate the parameters $x^{(j)}$, $y^{(j)}$ and $z^{(j)}$ to Equation \ref{eqn:StochasticPayoff}, we need to compute the expected value $\mathbb{E}\left(\Pi^{(j)}\right)$. In the form given, this may be complex, since the functions defined in Equation \ref{eqn:SinglePayOff} maybe non-linear, meaning we cannot pass the expectation operator through the expression.

The solution to the game is then defined by the simultaneous optimization problem:
\begin{equation}
\forall j \left\{
\begin{aligned}
\max \;\; & \mathbb{E}\left(\Pi^{(j)}(\mathbf{x}(t),\mathbf{y}(t),\mathbf{z}(t))\right)\\
s.t. \;\; & x^{(j)}(t) + y^{(j)}(t) + z^{(j)}(t) = 1 \quad \forall t\\
& x^{(j)}(t), y^{(j)}(t), z^{(j)}(t) \geq 0 \quad \forall t
\end{aligned}\right.
\end{equation}
where $\mathbf{x}(t)$, $\mathbf{y}(t)$, $\mathbf{z}(t)$ are the vectors of decision variables for the players. Let 
\begin{multline}
\Omega = \prod_{j,t} \left\{(x^{(j)}(t),y^{(j)}(t),z^{(j)}(t)) \in [0,1]^3 :\right.\\ 
x^{(j)}(t) + y^{(j)}(t) + z^{(j)}(t) = 1, \\
\left.x^{(j)}(t), y^{(j)}(t), z^{(j)}(t) \geq 0 \right\}
\end{multline}
This is the \textit{complete} strategy space for all players over the course of time $t \in [0,T]$. Any Nash equilibrium will be chosen from this strategy space. Theorem 1 of \cite{Ros63} provides the following (uninteresting) result:

\begin{proposition} Suppose that $\mathbb{E}\left(\Pi^{(j)}(\mathbf{x}(t),\mathbf{y}(t),\mathbf{z}(t))\right)$ is concave for all $j$ then there is a Nash equilibrium in $\Omega$ for this game.   
\end{proposition}

\begin{remark}
The uniqueness of a Nash equilibrium in this case is completely a function of the structure of specific objective functions.
\end{remark}

We noted above that the structure of $\mathbb{E}\left(\Pi^{(j)}(\mathbf{x}(t),\mathbf{y}(t),\mathbf{z}(t))\right)$ maybe complex. By way of simplification, we can write a specific form of Equation  \ref{eqn:SinglePayOff}, as: 
\begin{multline}
\Pi^{(j)}(t) = \\ 
w_1\alpha(x^{(j)},z^{(j)},\bar{x}^{(j)},\bar{z}^{(j)},t)\left(X^{(j)}(t)+(1-\tau(t))Z^{(j)}(t)\right) \\+ 
w_2\left(\beta(y^{(j)},\bar{y}^{(j)},t)Y^{(j)}(t) + \eta Z^{(j)}(t)\right)  \\
-w_3\gamma(z^{(j)},\bar{z}^{(j)},t)\tau(t)Z^{(j)}(t) - w_4\zeta Z^{(j)}(t) \\ 
-w_5\theta X^{(i)}(t)
\end{multline}
Here we replace the functions from Equation \ref{eqn:SinglePayOff} with piecewise constant multipliers. We can then relate Equation \ref{eqn:StochasticPayoff} to the parameters $x^{(j)}$, $y^{(j)}$ and $z^{(j)}$, we note that:
\begin{multline}
\mathbb{E}\left(\Pi^{(j)}\right) = 
\sum_{t = 0}^{T}\rho^t \cdot \\ \left[ 
w_1\alpha(x^{(j)},z^{(j)},\bar{x}^{(j)},\bar{z}^{(j)},t)\left(x^{(j)}(t)+(1-\tau(t))z^{(j)}(t)\right)  \right.\\
+w_2\left(\beta(y^{(j)},\bar{y}^{(j)},t)y^{(j)}(t) + \eta z^{(j)}(t)\right) - \\
w_3\gamma(z^{(j)},\bar{z}^{(j)},t)\tau(t)z^{(j)}(t) - \\
\left. w_4\zeta z^{(j)}(t) - w_5\theta x^{(i)}(t)
\right]
\label{eqn:ExpectedPayoff}
\end{multline}
That is, a user's expected payoff after engaging in this game, is a function of the proportion of time he releases information, withholds information and lies about information. Moreover, because we assume the  reward/cost multipliers ($\alpha$, $\beta$ and $\gamma$) are dependent on the group average rates of releasing, withholding and lying about information, the payoff to Player $j$ is dependent on the choices of all other players in his circle of friends. Thus, a game dynamic is established. We study this simplified game form in the remainder of the paper.

\section{Evolutionary Dynamics and Example}
\label{sec:EvolutionaryGame}
A critical problem with the game defined in the previos section is that individuals will never optimize their behavior according to it. An individual can estimate many of the parameters in the model, but will never make decisions based on the long run objective of maximizing his utility function.  It will simply be impossible for an individual to chose an optimizing strategy \textit{ab initio}, particularly without having a clear understanding of the strategies of other players. This is especially true if (as is likely the case) multiple Nash equilibria exist.

However, a user may engage in a more evolutionary model of decision making \cite{Wei95}\footnote{Weibull's book, \cite{Wei95} is an introduction to evolutionary game theory, which inspires the approach described herein. We are not proposing a classical evolutionary game. Instead, we are proposing an evolutionary mechanism applied to a game theoretic context that describes user learning.}:
\begin{enumerate*}
\item At any time $t$, Player $j$ has a strategy $(x^{(j)}(t),y^{(j)}(t),z^{(j)}(t))$ with an initial strategy $(x^{(j)}(0),y^{(j)}(0),z^{(j)}(0))$

\item At each time $t$, Player $j$ will solve the one-stage game derived from Equation \ref{eqn:ExpectedPayoff} by finding a maximizing strategy $\left(\hat{x}^{(j)}, \hat{y}^{(j)}, \hat{z}^{(j)}\right)$ with respect to the current observed strategies of the other players and the current parameters in the model.

\item Each player's strategy is updated by the rule:
\begin{gather}
x^{(j)}(t+1) = x^{(j)}(t) + \epsilon^{(j)}\left(\hat{x}^{(j)} - x^{(j)}(t)\right)\label{eqn:Dyn1}\\
y^{(j)}(t+1) = y^{(j)}(t) + \epsilon^{(j)}\left(\hat{y}^{(j)} - y^{(j)}(t)\right)\\
x^{(j)}(t+1) = z^{(j)}(t) + \epsilon^{(j)}\left(\hat{z}^{(j)} - z^{(j)}(t)\right)\label{eqn:Dyn3}
\end{gather}
\end{enumerate*}
Here $\epsilon^{(j)}$ is a learning rate associated to the player, and is assumed to be small -- that is, $\epsilon^{(j)} \ll 1$.  The dynamics given in Equations \ref{eqn:Dyn1} - \ref{eqn:Dyn3} are a discrete variation of the Jacobi iteration for finding equilibria in games (see e.g., \cite{JK02}). At its core, this is just a form of gradient ascent.

Intuitively, each time a user makes a decision about a piece of information, he considers his knowledge of the other players and computes an optimal move for this time period. However, instead of changing his strategy completely, he modifies his strategy (learns) by a small amount in the direction of optimality. This is consistent with an individual who learns the average behavior of the social network.

\subsection{Example}
By way of example, assume we have a small clique of three friends on a social network (this is the graph governing Equation \ref{eqn:Average}). Consider the following multiplier definitions for use in Equation \ref{eqn:ExpectedPayoff}. These functions are  derived from  a   qualitative analysis of the data collected in the experiment described in the previous sections,  and  are intended to be simple token counting margin functions.
\begin{equation}
\alpha(x^{(j)},z^{(j)},\bar{x}^{(j)},\bar{z}^{(j)}) = \begin{cases}
1 & x^{(j)}+z^{(j)} \leq \bar{x}^{(j)}+\bar{z}^{(j)}\\
0 & \text{otherwise}
\end{cases}
\end{equation}
In this case, the social value of information is non-zero only if the information provided is in some way lower than the mean information provided by the group. That is to say, you can social capital only if you're not posting more information than the group average. However, if the group average is high, you will accrue social capital the more you post. 

\begin{equation}
\beta(y^{(j)},\bar{y}^{(j)}) = \begin{cases}
1 & y^{(j)}\geq \bar{y}^{(j)}\\
0 & \text{otherwise}
\end{cases}
\end{equation}
Here, the privacy value is non-zero only if the amount of information that is to be admitted is larger than average amount of information being admitted. Similarly, we can define:
\begin{equation}
\gamma(z^{(j)},\bar{z}^{(j)}) = \begin{cases}
1 & z^{(j)}\geq \bar{z}^{(j)}\\
0 & \text{otherwise}
\end{cases}
\end{equation}
In this case, the cost of a lie is only non-zero if the lie is in some sense more egregious than the average level of dishonesty. Finally, we can set: $\zeta=1$ and $\theta=1$.

Variations in the players behavior can then be created by modifying $w_1,\dots,w_5$ in Equation \ref{eqn:ExpectedPayoff}. Finally, we assume that $\tau$ increases linearly in time from $0.1$ to $0.9$. Recall, $\tau$ is the probability that a lie can be detected by the social network. Thus, as time proceeds, it becomes more likely that a falsehood is detected because more information is available about each player \footnote{In a  fully formalized model, we believe that $\tau$ will be a function of $Q$ (defined in Equation \ref{eqn:Q}), however for our simple studies, this is sufficient}.

\begin{figure}[htbp]
\centering
\includegraphics[scale=0.29]{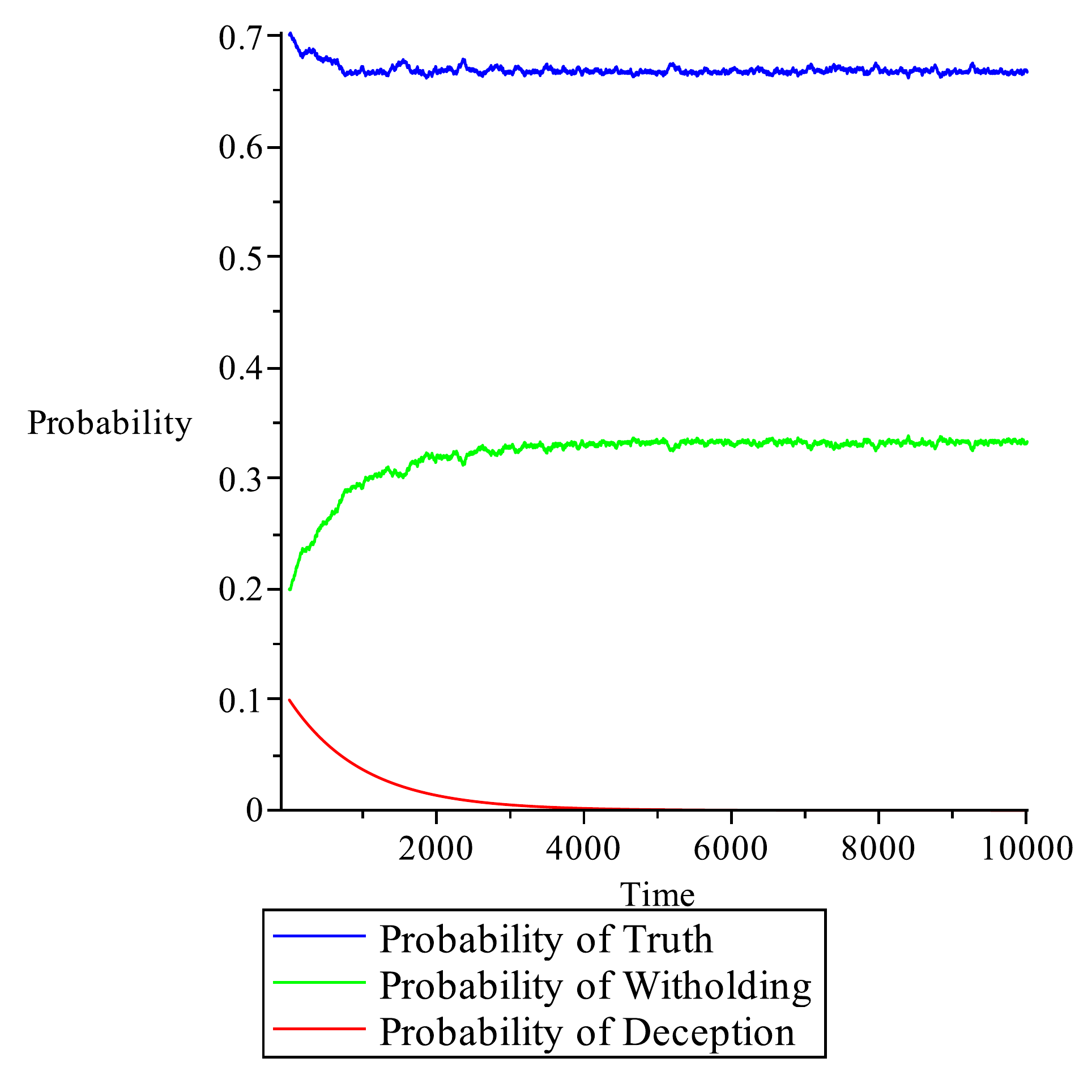}
\caption{Evolutionary output of a game with three identical players in which $w_1 = 1$, $w_2 = 0.25$, $w_3 = 0.5$, $w_4 = 1$ and $w_5 = 0.125$, suggesting that social capital is much more important than the gain associated with privacy.}
\label{fig:Game1}
\end{figure}

We study three specific examples of the dynamics produced by this model to illustrate that even {\em  these (simple) dynamics are capable of qualitatively reproducing behavior observed in our study}. In the first example, social capital is deemed more important than privacy ($w_1 = 1$, $w_2 = 0.25$ and $w_5 = 0.125$), but there is a stronger sense of morality ($w_3 = 0.5$, $w_4 = 1$). We assume three identical players connected by a complete graph with three vertices. Player evolution is illustrated in Figure \ref{fig:Game1}
When we start the game with three players, each playing the strategy $x^{(j)}=0.7$, $y^{(j)} = 0.2$ and $z^{(j)} = 0.1$, we see deception is removed from the system relatively quickly, while information hiding increases (replacing deception in the system). Notice the system converges to a stationary strategy near $x^{(j)}=2/3$, $y^{(j)} = 1/3$ and $z^{(j)} = 0$. In simpler terms, these equilibrium points are consistent with our findings that social capital is much more important than the gain associated with privacy, because we see that there is a preference toward sharing information truthfully. This was observed in Section \ref{factors}. It is also worth noting, that it can be shown numerically these are limiting Nash equilibria for this example.

\begin{figure}[htbp]
\centering
\includegraphics[scale=0.29]{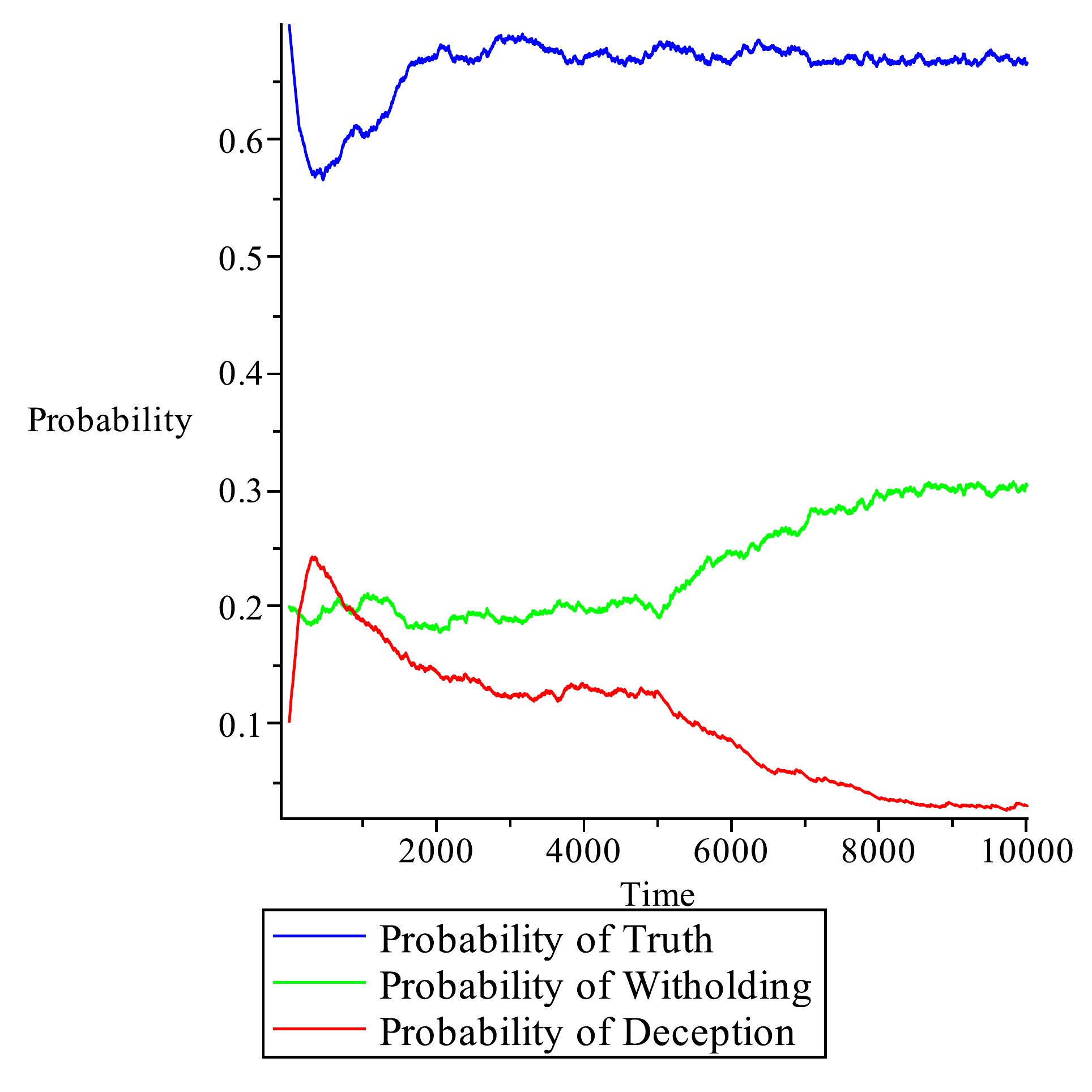}
\caption{Evolutionary output of a game with three identical players in which $w_1 = 2$, $w_2 = 0.25$, $w_3 = 0.25$, $w_4 = 0.125$ and $w_5 = 0.125$, suggesting that social capital is much more important than the gain associated with privacy and morality is of little concern.}
\label{fig:Game2}
\end{figure}

By way of comparison, we can construct a game with less morality and even more importance associated to social capital (being obtained by any means necessary) with $w_1 = 2$, $w_2 = 0.25$, $w_3 = 0.25$, $w_4 = 0.125$ and $w_5 = 0.125$. Note, $w_4 = 0.125$ indicates a low moral penalty for lying. The evolution of the players is illustrated in Figure \ref{fig:Game2}.

The dynamics in this game are different, than those of the first game. There is an initial, substantial, increase in the level of deception (when it is easy to lie) in order to obtain social capital. As it becomes more difficult to lie, the players return to a more truthful scenario that is easier to support. 
This example confirms the identified effects of the signaling theory in our dataset: users are less likely to deceive  when they are heavily involved in social interactions (Section III-C1 and III-C2).

\begin{figure}[htbp]
\centering
\includegraphics[scale=0.29]{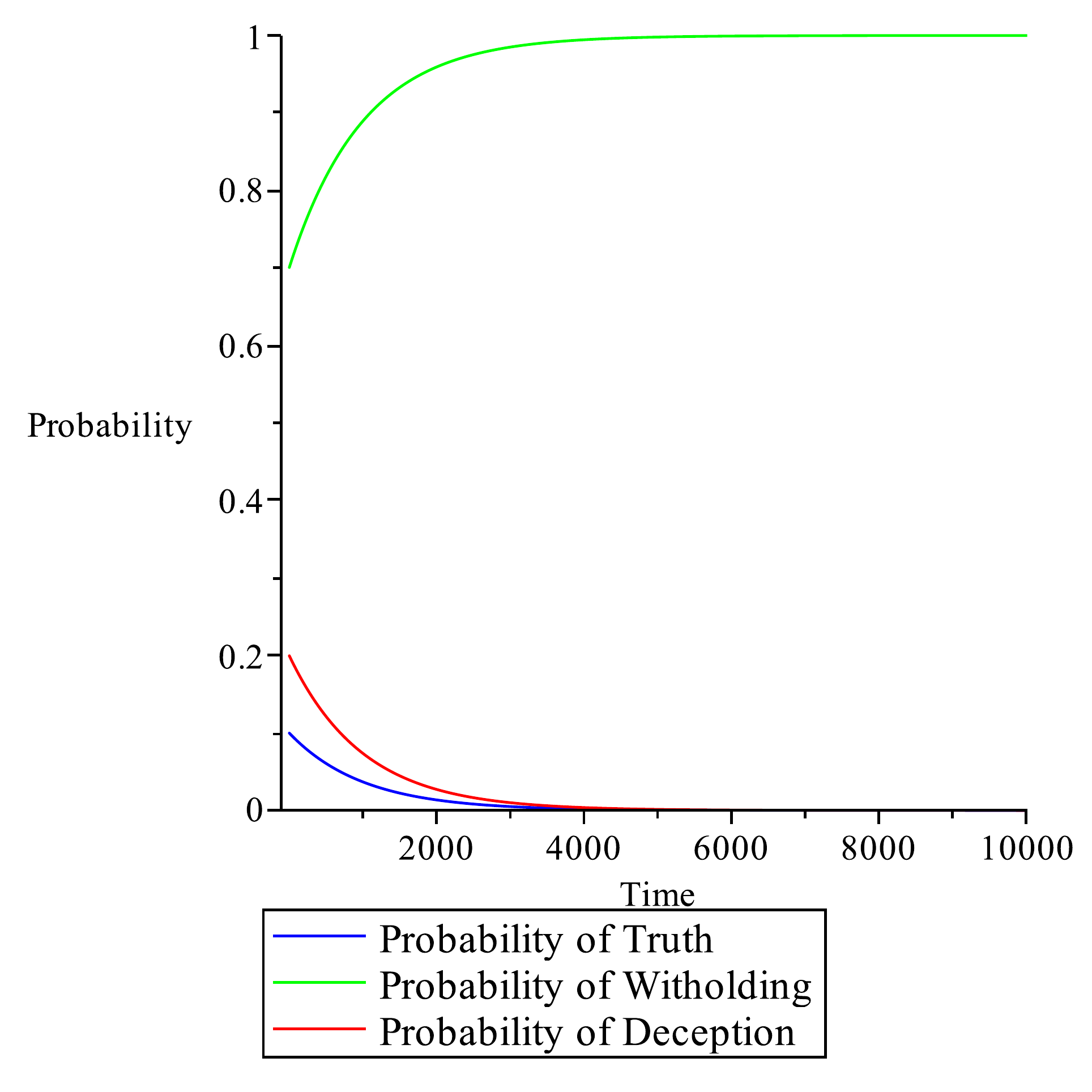}
\caption{Evolutionary output of a game with three identical players in which $w_1 = 0.5$, $w_2 = 5$, $w_3 = 2$, $w_4 = 100$ and $w_5 = 3$, suggesting that morality and privacy are paramount to this user.}
\label{fig:Game3}
\end{figure}
In our final example, we consider a \textit{highly} moral player who puts less emphasis on social capital and substantial emphasis on information privacy. In this case, we have: $w_1 = 0.5$, $w_2 = 5$, $w_3 = 2$, $w_4 = 100$ and $w_5 = 3$. The results are illustrated in Figure \ref{fig:Game3}.
This objective function models a user who is highly moral deciding whether to release an information type that may be sensitive, such as GPA or dating status and illustrates the ability of the model to capture the various qualitative results observed in the survey. In particular, this result is consistent with the finding reported in Section \ref{sec:moral}: withholding  is used as a form of control when deception is considered unethical. 

The proposed approach can also be used to study richer scenarios, including those with players that  begin with different strategies and games that are played on distinct graph structures as was discussed in \cite{GS12}.

\section{Future Directions} \label{sec:conc}
In this work, we have shown an informed model on deception and misrepresentation in OSNs.  The model is derived from a token counting approach modeled in a stochastic, continuous Petri net. This net is then used to derive an objective function for each player in which the payoff to a given player is a function not only of his decisions but also the decisions of his circle of friends. This leads to a game theoretic framework. We show that while this game has at least one Nash equilibria, it is more interesting to consider an evolutionary game dynamic in which players learn over time and converge to a stationary strategy. 

We are left with a number of unsolved questions, that we plan to explore in the near future.  First, we are interested in collecting more detailed data from  real-world users, to deepen our understanding of users' interactions and identity revelation processes.  For example, in the current study we did not focus on the users' actions, that result in  identity disclosure.  What are the typical passive social transactions (post an item on your page which may be ÒsilentlyÓ consumed by those who've been given access to it) or active transactions  (sharing, commenting on other's  content or status updates, give feedback) that lead to information revelation and/or to deception?  How do different ÒoutcomesÓ of such transactions affect social capital, and therefore result in  truthful and untruthful information sharing? How do secondary (friend-of-friend, triad) relationships influence information sharing? 
Results obtained from these studies will guide the next step of our research on the model. 
%

For example, using this information, we would like to determine the structural characteristics of the benefit and cost functions in Equation \ref{eqn:StochasticPayoff}. In addition to this, it would be useful to identify whether the dynamics described force the players to converge to an equilibria of some type. We expect that convergence to an equilibrium point should occur, but it is not clear if this is a global property of all well-behaved payoff functions.  As noted in \cite{GS12} there can be an interaction between the properties of the graph on which the game is played and the number and type of symmetric equilibria. We have not explored this using the model presented in this paper, but we believe this is a necessary step in understanding the behavior of user dynamics in information expression in social networks. 
\section*{Ackowledgement}
Portions of Dr. Griffin's work were supported by the Army Research Office under Grant W911NF-11-1-0487.

\bibliographystyle{abbrv}
\bibliography{biblio}

\end{document}